\documentclass[aps,pre,onecolumn,epsf,showpacs,graphicx]{revtex4}
\usepackage{epsfig,latexsym} 
\usepackage{amsmath,amscd}

\begin{document}

\title{\bf\noindent Extreme Value  Statistics of Eigenvalues of Gaussian Random Matrices}
\author{David S. Dean$^1$ and Satya N. Majumdar$^2$}
\affiliation{
$^1$ Laboratoire de Physique Th\'eorique (UMR 5152 du CNRS),
Universit\'e Paul Sabatier, 118, route de Narbonne, 31062
Toulouse Cedex 4, France\\
$^2$ Laboratoire de Physique Th\'eorique et
Mod\`eles Statistiques (UMR 8626 du CNRS),
Universit\'e Paris-Sud, B\^at. 100, 91405 Orsay Cedex, France}


\begin{abstract}
We compute exact asymptotic results for the probability of the occurrence of large 
deviations
of the largest (smallest) eigenvalue of random matrices belonging to the 
Gaussian orthogonal, unitary and symplectic ensembles. In particular, we show
that the probability that all the eigenvalues of an $(N\times N)$
random matrix are positive (negative) decreases for large $N$ 
as $\sim \exp[-\beta \theta(0) N^2]$
where the Dyson index $\beta$ characterizes the ensemble and the exponent
$\theta(0)=(\ln 3)/4=0.274653\dots$ is universal. We compute 
the probability that the eigenvalues lie in the interval 
$[\zeta_1,\zeta_2]$ which allows us to calculate the
joint probability distribution of the minimum and the 
maximum eigenvalue. As a byproduct, we also obtain
exactly the average density of states in Gaussian ensembles
whose eigenvalues  
are restricted 
to lie in the interval $[\zeta_1,\zeta_2]$, 
thus generalizing the celebrated 
Wigner semi-circle law to these restricted ensembles. 
It is found that the density of states generically exhibits an inverse 
square-root singularity at the location of the barriers. 
These results are confirmed 
by numerical simulations.

\end{abstract}  
\maketitle

\section{Introduction}
Studies of the statistics of the eigenvalues of random matrices have a long 
history going back to the seminal work of Wigner \cite{wig}.  Random matrix 
theory has been successfully applied in various branches of physics and mathematics,
including in subjects ranging from nuclear physics, quantum chaos, 
disordered systems, string theory and even in number theory~\cite{mehta}.
Of particular importance are Gaussian random matrices whose entries
are independent Gaussian variables~\cite{mehta}.
Depending on the physical symmetries of the problem, three classes of 
matrices with Gaussian entries arise~\cite{mehta}: $(N\times
N)$ real symmetric (Gaussian Orthogonal Ensemble (GOE)), $(N\times N)$
complex Hermitian (Gaussian Unitary Ensemble (GUE)) and $(2N\times
2N)$ self-dual Hermitian matrices (Gaussian Symplectic Ensemble
(GSE)). In these models the probability distribution for a matrix 
$M$ in the ensemble is given by
\begin{equation}
p(M) \propto \exp\left(-\frac{\beta}{2}(M,M)\right),
\end{equation}
where $(M,M)$ is the inner product on the space of matrices invariant,
under orthogonal, unitary and symplectic transformations respectively
and the parameter $\beta$ is the Dyson index.
In these three cases the inner products and the Dyson indices are given by
\begin{eqnarray}
  (M,M) &=& \rm{Tr}(M^2); \quad \beta=1 \quad\quad \ \ \ \   {\rm GOE}    \\
  (M,M) &=& \rm{Tr}(M^* M); \quad \beta=2 \quad\quad  {\rm GUE}   \\
  (M,M) &=& \rm{Tr}(M^\dagger M);\quad \beta=4\quad\quad     {\rm GSE} 
\end{eqnarray}
where $\cdot^*$ denotes the hermitian conjugate of complex valued
matrices and $\cdot^\dagger$ denotes the symplectic conjugate on
quaternion valued matrices. The above quadratic actions are the
simplest forms (corresponding to free fields) of matrix models which
have been extensively studied in the context of particle physics and
field theory.

\begin{figure}
%\centerline{\epsfxsize\columnwidth \epsfbox{tw.eps}}
\includegraphics[width=.45\textwidth]{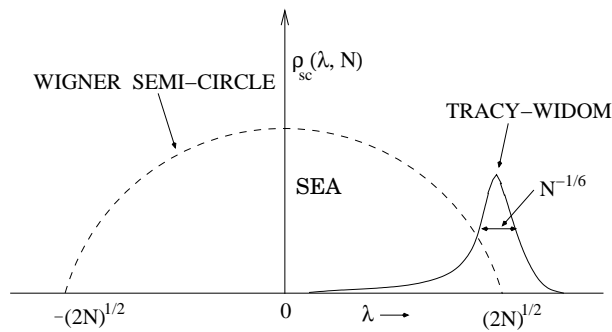}
\caption{The dashed line shows the Wigner semi-circular form of the
average density of states. The largest eigenvalue is centered around its mean $\sqrt{2N}$
and fluctuates over a scale of width $N^{-1/6}$. The probability of fluctuations
on this scale is described by the Tracy-Widom distribution (shown schematically).}
\label{figtw}
\end{figure}

A central result in the theory of random matrices is the
celebrated Wigner semi-circle law. It states that for large $N$ and
{\em on an average}, the $N$ eigenvalues lie within a
finite interval $\left[-\sqrt{2N}, \sqrt{2N}\right]$, often referred
to as the Wigner `sea'. Within this sea, the average density of states
has a semi-circular form (see Fig.( \ref{figtw})) that vanishes at the
two edges $-\sqrt{2N}$ and $\sqrt{2N}$
\begin{equation}
\rho_{\rm sc}(\lambda,N) = \sqrt{\frac{2}{N\pi^2}}\,
{\left[1 -\frac{\lambda^2}{2N}\right]}^{1/2}.
\label{wig1}
\end{equation}
The above result means that, if one looks at the density of states of a
typical system described by one of the three ensembles above, for a
large enough system, it will resemble closely the Wigner semi-circle law.

While the semi-circle law provides a global information about
how the eigenvalues are {\em typically} distributed, unfortunately it does not
contain enough information about the probabilities of {\em rare} events. 
The questions concerning rare events have recently come up in different contexts.
For example, string theorists have recently been confronted by the possibility that
there may be a huge number of effective theories describing our
universe.  The landscape made up of these theories is called the
string landscape. This seemingly embarrassing situation may however
help to explain certain fine tuning puzzles in particle physics. The
basic argument is as follows.  Our universe is one which supports
intelligent life and this requires that the ratios of certain
fundamental constants lie in specific ranges. The fine tuning we
observe is thus a necessary condition that we are there to describe
it. Other possible universes would have different vacua but there
would be no intelligent life to study them. This approach to string
theory is called the anthropic principle based string theory and is a
subject of current and intense debate.  In \cite{string,AE} the authors
carried out an analysis of the string landscape based solely on the
basis that it is described by a large $N$ multi-component scalar
potential.  Of particular interest is the determination of the typical
properties of the vacua in the string landscape based only on
assumptions about the dimensionality of the landscape and other simple
general features. The motivation for these studies is to determine to
what extent the string landscape is determined by large $N$ statistics
and what features depend on the actual structure of the underlying string theory.

One of the main questions posed in \cite{AE} is: for a
$(N\times N)$ Gaussian random matrix, what is
the probability $P_N$ that all its eigenvalues are positive (or negative)
\begin{equation}
P_N= {\rm Prob}[\lambda_1\ge 0, \lambda_2\ge 0,\ldots, \lambda_N\ge 0] ?
\label{pn1}
\end{equation}
From the semi-circle law, we know that on an average half the eigenvalues are positive 
and half of them are negative. Thus the event that all eigenvalues are
negative (or positive) is clearly an atypical rare event.
This question arises in the so called `counting problem'
of local minima in a random multi-field potential or a 
landscape. Given a stationary point of the landscape,
if all eigenvalues of the Hessian matrix of the potential
are positive, clearly the stationary point is a local
minimum. Thus, the probability that all eigenvalues
of a random Hessian matrix are positive provides as estimate for the
fraction of local minima amongst the stationary points
of the landscape. In particular, the authors of \cite{AE}
studied the case where the Hessian matrix was
drawn from a GOE ensemble ($\beta=1$). Thus, in this
context $P_N$ is just the probability that a random GOE
matrix is positive definite. This probability has
also been studied in the mathematics literature~\cite{DedMal}
and one can easily compute $P_N$ for smaller values of $N=1,2,3$.
For example, one can show that~\cite{DedMal}
\begin{equation}
P_1=1/2,\quad P_2=\frac{2-\sqrt{2}}{4}, \quad P_3=\frac{\pi-2\sqrt{2}}{4\pi}. 
\label{PSN}
\end{equation}
The interesting question is how $P_N$ behaves for large $N$?
It was argued in \cite{AE} that for large $N$, $P_N$ decays
as $P_N\sim \exp\left(-\theta(0)N^2\right)$ where the decay constant
$\theta(0)$ was estimated to be $\approx 1/4$ numerically and via a heuristic
argument. The scaling of this probability with $N^2$ is not surprising
and has been alluded to in the literature for a number of years,
notably in relation to studies of the distribution of the index (the
number of negative eigenvalues) of Gaussian matrices~\cite{index,Fyod1}. 
However an exact expression for $\theta(0)$ was not available
until only recently, when the short
form of this paper \cite{letter} was published. In \cite{letter}
we had shown that for all the three Gaussian ensembles, to leading
order in large $N$
\begin{equation}
P_N \sim \exp[-\beta \theta(0) N^2]; \quad\, {\rm where}\quad\, 
\theta(0)= \frac{\ln 3}{4}=(0.274653\ldots). 
\label{theta0}
\end{equation}
Interestingly, the probability $P_N$ has also recently shown up
in a rather different problem in mathematics. Dedieu and Maljovich
has shown~\cite{DedMal} recently that $P_N$ is exactly equal to the expected 
number of minima of a random polynomial of degree at most $2$
and $N$ variables. Our result in \cite{letter} thus provided
an exact answer to this problem for large $N$~\cite{DedMal}. 

To put our results in a more general context, we note that the semi-circle law
tells us that the average 
 of the maximum (minimum) eigenvalue is $\sqrt{2N}$
(-$\sqrt{2N}$).  However, for finite but large $N$, the maximum
eigenvalue fluctuates, around its mean $\sqrt{2N}$, from one sample to
another. Relatively recently Tracy and Widom~\cite{TW} proved that
these fluctuations {\em typically} occur over a narrow scale of $\sim
O(N^{-1/6})$ around the upper edge $\sqrt{2N}$ of the Wigner sea (see
Fig. 1). More precisely, they showed~\cite{TW} that asymptotically for
large $N$, the scaling variable $\xi=\sqrt{2}\,N^{1/6}\, [\lambda_{\rm
max}-\sqrt{2N}]$ has a limiting $N$-independent probability
distribution, ${\rm Prob}[\xi\le x]= F_{\beta}(x)$ whose form depends
on the value of the parameter $\beta=1$, $2$ and $4$ characterizing
respectively the GOE, GUE and GSE. The function $F_{\beta}(x)$,
computed as a solution of a nonlinear differential equation~\cite{TW},
approaches to $1$ as $x\to \infty$ and decays rapidly to zero as $x\to
-\infty$. For example, for $\beta=2$, $F_2(x)$ has the following
tails~\cite{TW},
\begin{eqnarray}
F_2(x) &\to & 1- O\left(\exp[-4x^{3/2}/3]\right)\quad\, {\rm as}\,\,\, x\to \infty
\nonumber \\
&\to & \exp[-|x|^3/12] \quad\, {\rm as}\,\,\, x\to -\infty.
\label{asymp1}
\end{eqnarray}
The probability density function $dF_{\beta}/dx$ thus has highly
asymmetric tails.  The distribution of the minimum eigenvalue simply
follows from the fact that ${\rm Prob}[\lambda_{\rm min}\ge
\zeta]={\rm Prob}[\lambda_{\rm max}\le -\zeta]$.  Amazingly, the
Tracy-Widom distribution has since emerged in a number of seemingly
unrelated problems~\cite{leshouches} such as the longest increasing subsequence
problem~\cite{BDJ}, directed polymers in $(1+1)$-dimensions~\cite{DP},
various $(1+1)$-dimensional growth models~\cite{Growth},
a class of sequence alignment problems~\cite{MN}, mesocopic fluctuations
in dity metal grains and semiconductor quantum dots~\cite{meso} and 
also in finance~\cite{BBP}.

The Tracy-Widom distribution describes the probability of {\em typical and small}
fluctuations of $\lambda_{\rm max}$ over a very narrow region of width
$\sim O(N^{-1/6})$ around the mean $\langle \lambda_{\rm max}\rangle
\approx \sqrt{2N}$. A natural question is how to describe the
probability of {\em atypical and large} fluctuations of $\lambda_{max}$ around its
mean, say over a wider region of width $\sim O(N^{1/2})$?
For example, the probability $P_N$ that all eigenvalues are negative (or positive) 
is the same as the probability that $\lambda_{\rm max}\le 0$
(or $\lambda_{\rm min}\ge 0$). Since $\langle \lambda_{\rm max}\rangle
\approx \sqrt{2N} $, this requires the computation of the probability
of an extremely rare event characterizing a large deviation of $\sim
-O(N^{1/2})$ to the left of the mean. In \cite{letter} we calculated the
exact large deviation function associated with large fluctuations of
$\sim -O(N^{1/2})$ of $\lambda_{\rm max}$ to the left of its
mean value $\sqrt{2N}$. It was shown that for large $N$ and for all ensembles
\begin{equation}
{\rm Prob}\left[\lambda_{\rm max}\le t, N\right] \sim \exp\left[-\beta
N^2 \Phi\left( \frac{\sqrt{2N}-t}{\sqrt{N}} \right) \right]
\label{ldf0}
\end{equation}
where $t\sim O(N^{1/2})\le \sqrt{2N}$ is located deep inside the
Wigner sea. The large deviation function $\Phi(y)$ is zero for $y\le 0$,
but is nontrivial for $y> 0$ which was computed exactly in \cite{letter}.
For {\em small} deviations to the left of the mean, taking the
$y\to 0$ limit of $\Phi(y)$, one recovers the left tail of the Tracy-Widom distribution
as in Eq. (\ref{asymp1}). Thus our result for {\em large} deviations
of $\sim -O(N^{1/2})$ to the left of the mean is complementary to the
Tracy-Widom result for {\em small} fluctuations of $\sim -O(N^{-1/6})$
and the two solutions match smoothly. Also, the probability $P_N$  
that all eigenvalues are negative (or positive) simply follows
from the general result in Eq. (\ref{ldf0}) by putting $t=0$,
\begin{equation}
P_N = {\rm Prob}\left[\lambda_{\rm max}\le 0, N\right]\sim \exp\left[-\beta 
\Phi\left(\sqrt{2}\right)\,N^2\right]
\label{pn2}
\end{equation}
thus identifying $\theta(0)=\Phi\left(\sqrt{2}\right)=(\ln 3)/4$, a special case of the general large 
deviation function. 

The purpose of this paper 
is to provide a detailed derivation of the above results announced in \cite{letter}, as
well as numerical results in support of our analytical formulas. In addition, 
we also derive asymptotic results for the joint probability distribution 
of the minimum and the maximum eigenvalue.

Statistical analysis motivated by anthropic considerations in string
theory or random polynomials in mathematics may seem a long way from
laboratory based physics, however similar questions appear naturally
also in providing criteria of physical stability in dynamical systems
or ecosystems~\cite{May,KL}.  Near a fixed point of a dynamical
system, one can linearize the equations of motion and the eigenvalues
of the corresponding matrix associated with the linear equations
provide important informations about the stability of the fixed
point. For example, if all the eigenvalues are negative (or positive)
the fixed point is a stable (or unstable) one.  In this context,
another important question arises naturally. Suppose that the
dynamical system is close to stable (unstable) fixed point, i.e., all
the eigenvalues are negative (positive). Given this fact, one may
further want to know how these negative (positive) eigenvalues are
distributed. In other words, what is the average density of states of
the negative (or positive) eigenvalues given the fact that one is
close to a stable (or unstable) fixed point.  In this paper we will
calculate the density of states in this conditioned ensemble and we
will see that it is quite different to the Wigner semi-circle law.
   
Recently the problem of determining the stability of the critical
points of Gaussian random fields in large dimensional spaces was
analyzed \cite{brde,FSW}. The Hessian matrix in this case does not have
the statistics of a Gaussian ensemble and the probability that a
randomly chosen critical point is a minimum case be shown to decay as
$\exp\left(-N\,\psi\right)$ and the exponent $\psi$ can be explicitly
calculated in terms of the two point correlation function of the
Gaussian field. Thus the scaling in $N$ is quite different to the
random matrix case and this scaling obviously makes minima much more
likely and yields a more usual thermodynamic scaling of the entropy of
critical points \cite{brde}. The statistics of the Gaussian field
problem are perhaps more relevant to statistical landscape scenarios
in string theory.

The paper is organized as follows. In Section II we begin by recalling the 
Coulomb gas representation of the distribution of eigenvalues of 
Gaussian matrices and  show how our problem can be formulated by placing 
a hard wall constraint on the Coulomb gas. This is the key step in the method and 
the technique has since been applied to analyze the statistics of critical points 
of Gaussian random fields
\cite{brde} and also to study the probability of rare fluctuations 
of the 
maximal eigenvalue of 
Wishart random matrices \cite{vmb}.
We then show how in the large 
$N$ limit the problem can be solved using a saddle point computation of
a functional integral and we discuss the features of our
analytic results. In Section III, we extend this method
to compute the asymptotic joint probability distribution
of the minimum and the maximum eigenvalue. This requires studying
the Coulomb gas confined between two hard walls. In Section IV we  carry 
out some  
numerical work to confirm our predictions about the probability of extreme
deviations of the maximal eigenvalue.
We show how the Coulomb gas formulation can 
again be exploited even for numerical purposes. Finally we present our conclusions
in Section V. 

\section{The Coulomb Gas Formulation and the Probability of Rare Fluctuations}

The joint probability density function (pdf) of the eigenvalues of 
an $N\times N$ Gaussian matrix is given by the classic result of Wigner~\cite{wig,mehta} 
\begin{equation}
P(\lambda_1, \lambda_2,\dots, \lambda_N) = B_N \exp\left[-\frac{\beta}{2}\left(\sum_{i=1}^N\lambda_i^2
-\sum_{i\ne j}\ln(|\lambda_i-\lambda_j|)\right)\right],
\label{pdf}
\end{equation}
where $B_N$ normalizes the pdf and $\beta=1$, $2$ and $4$ correspond
respectively to the GOE, GUE and GSE.
The joint law in Eq. (\ref{pdf}) allows one to
interpret the eigenvalues as the positions of charged particles,
repelling each other via a $2$-d Coulomb potential (logarithmic);
they are confined on a $1$-d line and each is subject to an external harmonic
potential. The parameter $\beta$ that characterizes the type of
ensemble can then be interpreted as the inverse temperature.

Once the joint pdf is known explicitly, other statistical properties of a random matrix
can, in principle, be derived from this joint pdf. In practice, however
this is often a technically daunting task. For example, suppose we want to
compute the average density of states of the eigenvalues defined as
$\rho(\lambda,N)= \sum_{i=1}^N\langle
\delta(\lambda-\lambda_i)\rangle/N$, which counts the average number of
eigenvalues between $\lambda$ and $\lambda + d\lambda$ per unit length.
The angled bracket $\langle \rangle$ denotes an average over the joint pdf.
It then follows that $\rho(\lambda,N)$ is simply the marginal of the joint pdf,
i.e, we fix one of the eigenvalues (say the first one) at $\lambda$ and integrate the joint pdf
over the rest of the $(N-1)$ variables.
\begin{equation}
\rho(\lambda,N)=\frac{1}{N} \sum_{i=1}^N\langle
\delta(\lambda-\lambda_i)\rangle =\int_{-\infty}^{\infty}\prod_{i=2}^N d\lambda_i \,
P(\lambda,\lambda_2,\dots, \lambda_N).
\label{marginal}
\end{equation}
Wigner computed this marginal and showed~\cite{wig} that for large $N$ and for all $\beta$
it has the semi-circular form in Eq. (\ref{wig1}). 

Here we are interested in calculating the probability $
Q_N(\zeta)$ that all eigenvalues are greater
than some value $\zeta$. This probability is clearly also equal
to the cumulative probability that the 
minimum eigenvalue $\lambda_{\rm min}={\rm min}(\lambda_1, 
\lambda_2, \ldots, \lambda_N)$ is greater than $\zeta$, i.e.,
\begin{equation}
Q_N(\zeta)= {\rm Prob}\left[\lambda_1\ge \zeta, \lambda_2\ge \zeta,\ldots, \lambda_N\ge 
\zeta\right]={\rm Prob}[\lambda_{\rm min}\ge \zeta, N].
\label{qnz1}
\end{equation}
Since the Gaussian random matrix has the $x\to -x$ symmetry, it follows that  
the maximum eigenvalue $\lambda_{\rm max}$ has the same statistics as $-\lambda_{\rm min}$.
Hence, it follows that
\begin{equation}
{\rm Prob}[\lambda_{\rm max}\le t, N]= Q_N(\zeta=-t).
\label{maxmin1}
\end{equation}
Hence, knowing $Q_N(\zeta)$ will also allow us to compute the cumulative distribution
of the maximum. Also, note that the probability $P_N$ that all eigenvalues are
positive (or negative), as defined in the introduction, is simply
\begin{equation}
P_N = Q_N(0).
\label{pnqn0}
\end{equation}
In what follows, we will compute $Q_N(\zeta)$ in the scaling limit where
$\zeta\sim \sqrt{N}$ for large $N$. For this we will employ the saddle point
method in the framework of the Coulomb gas.

By definition 
\begin{equation}
Q_N(\zeta)= \int_{\zeta}^{\infty}\ldots\int_{\zeta}^{\infty} P(\lambda_1,\lambda_2,\ldots, 
\lambda_N)\,d\lambda_1\,d\lambda_2\ldots d\lambda_N
\label{qnz2}
\end{equation}
where $P$ is the joint pdf in Eq. (\ref{pdf}). This multiple integral can be written as 
a ratio
\begin{equation}
Q_N(\zeta)= \frac{Z_N(\zeta)}{Z_N(-\infty)}
\label{qnz3}
\end{equation}
where the partition function $Z_N(\zeta)$ is defined as
\begin{equation}
Z_N(\zeta)= \int_{\zeta}^{\infty}\ldots\int_{\zeta}^{\infty}
\, \exp\left[-\frac{\beta}{2}\left(\sum_{i=1}^N\lambda_i^2
-\sum_{i\ne j}\ln(|\lambda_i-\lambda_j|)\right)\right]\,d\lambda_1\,d\lambda_2\ldots d\lambda_N.
\label{pf1}
\end{equation}
Note that the normalization constant $B_N= 1/{Z_N(-\infty)}$. Clearly, for any finite 
$\zeta$,
$Z_N(\zeta)$ represents the partition function of the Coulomb gas which is
constrained in the region $[\zeta,\infty]$, i.e., it has a hard wall at
$\zeta$ ensuring that there are no eigenvalues to the left of $\zeta$. 

From the semi-circle law, it is evident a typical eigenvalue scales as $\sqrt{N}$
for large $N$. This is also evident from Eq. (\ref{pf1}) where the Coulomb interaction term
typically scales as $N^2$ for large $N$ while the energy corresponding to the external
potential scales as $\lambda^2 N$. In order that they balance, it follows that
typically $\lambda\sim \sqrt{N}$. It is therefore natural to rescale
$\mu_i = \lambda_i/\sqrt{N}$ so that the rescaled eigenvalue $\mu_i\sim O(1)$ for large $N$.
In terms of the rescaled eigenvalues, the partition function reads 
\begin{equation}
Z_N(\zeta) \propto \nonumber \int_{\mu_i >\frac{\zeta}{\sqrt{N}}}^\infty
\prod_{i=1}^N d\mu_i \exp\left(-\beta H({\bf \mu})\right),
\label{pf2}
\end{equation}
where the Hamiltonian of the Coulomb gas is 
\begin{equation}
H({\bf \mu}) = \frac{N}{2} \sum_{i=1}^N \mu_i^2 - \frac{1}{2}\,\sum_{i\neq j} \ln(
|\mu_i-\mu_j|).
\label{Hamil1}
\end{equation}

\subsection{Functional Integral, Large $N$ Saddle Point Analysis and the Constrained 
Charge Density}

We first define the normalized (to unity) spatial density field of the 
particles $\mu_i$ as
\begin{equation}
\rho(\mu) = \frac{1}{N}\,\sum_{i=1}^N \delta(\mu -\mu_i).
\end{equation}
This is just a `counting' function so that $\rho(\mu)\,d\mu$ counts the fraction of 
eigenvalues
between $\mu$ and $\mu+d\mu$. 
The energy of a configuration of the $\mu_i$'s can be expressed in terms
of the density $\rho$ as
\begin{equation}
H[\rho] =  {N^2} {\cal E}[\rho]
\end{equation}
where 
\begin{equation}
{\cal E}[\rho] = \frac{1}{2}\,\int d\mu\, \mu^2\, \rho(\mu) - \frac{1}{2}\,\int d\mu\, 
d\mu'
\rho(\mu)\, \rho(\mu')\, \ln(|\mu-\mu'|) +{1\over 2N}\int d\mu\ \rho(\mu)
\ln\left(l(\mu)\right).
\label{sigmaf1}
\end{equation}
The last term above removes the self interaction energy from the 
penultimate term and $l(\mu)$ represents a position dependent
cut-off. Dyson \cite{dyson} argued that $l(\mu) \sim 1/\rho(\mu)$
and hence this correction term has the form
\begin{equation}
\int d\mu\ \rho(\mu)
\ln\left(l(\mu)\right) = -\int d\mu\ \rho(\mu)
\ln\left(\rho(\mu)\right) + C',
\end{equation}
where $C'$ is a constant that cannot be determined by this argument, but  
one may assume that it is independent of the position of the hard wall. However,
in this paper, we will be interested only in the leading $O(N^2)$ behavior
and hence precise value of the constant $C'$ does not matter. 

%as we are only interested in the leading $O(N^2)$ behavior the relevant 
%action to this order is given by
%\begin{equation}
%\Sigma_0[\rho] = \frac{1}{2}\,\int d\mu\, \mu^2\, \rho(\mu) - \frac{1}{2}\,\int d\mu\, 
%d\mu'
%\rho(\mu)\, \rho(\mu')\, \ln(|\mu-\mu'|).
%\label{sigmaf}
%\end{equation}
%xxxx

The hard wall constraint can now be implemented simply by the condition
\begin{equation}
\rho(\mu) = 0\ {\rm for} \ \mu < {\frac{\zeta}{\sqrt{N}}}.
\end{equation}
The partition function may now be written as a functional integral over
the density field $\rho$ as
\begin{equation}
Z_N(\zeta) = \int d[\rho] \ J[\rho] \exp\left(-\beta N^2 {\cal E}[\rho]\right),
\end{equation} 
where $J[\rho]$ is the Jacobian involved in changing from the coordinates
$\mu_i$ to the density field $\rho$. Physically this Jacobian takes into account
the entropy associated with the density field $\rho$. For the sake of completeness we
will re-derive a familiar form of the Jacobian $J[\rho]$. Clearly $J$ can be written, up
to a constant prefactor $D_N$, as

\begin{equation}
J[\rho] = D_N\int \prod_{i=1} d\mu_i \ \delta\left[ N\rho(\mu) - \sum_i 
\delta(\mu-\mu_i)\right]
\end{equation}

where the integration range of the $\mu_i$ above are restricted to
the appropriate region. One now proceeds by making a functional
Fourier transform representation of the delta function at each point
$\mu$. This gives
\begin{equation}
J[\rho] = D'_N\int \prod_{i=1} d\mu_i d[g]\ \exp\left[\int d\mu 
\ g(\mu) \left[N\rho(\mu) - \sum_i \delta(\mu-\mu_i)\right]\right],
\end{equation}
where each $g(\mu)$ integral is along the imaginary axis and $D'_N$ is
a constant prefactor. The integral
over the $\mu_i$ may now be carried out giving
\begin{eqnarray}
J[\rho] &=& D'_N\int  d[g]\ \exp\left[N\,\int d\mu 
\ g(\mu)\,\rho(\mu)\right]\,\prod_{i=1}^N d\mu_i \exp\left[-g(\mu_i)\right] \nonumber \\
&=& D'_N\int  d[g]\ \exp\left[N\,\int d\mu
\ g(\mu)\,\rho(\mu) +   N\ln\left(\int 
d\mu \exp(-g(\mu))\right) \right].
\end{eqnarray}
The above functional integral over $g$ can be evaluated  by saddle point 
for large  $N$ and the corresponding saddle point equation (obtained via 
stationarity with respect to $g$) is 
\begin{equation}
 \rho(\mu) = \frac{\exp\left(-g(\mu)\right)}{ \int d\mu' \exp(-g(\mu'))}.
\end{equation}
We see that the normalization 
\begin{equation}
\int d\mu\ \rho(\mu) = 1 
\label{normf}
\end{equation}
is respected. Substituting in this solution for $g$ we find that
\begin{equation}
J[\rho] = D'_N \exp\left[- N\int d\mu \rho(\mu)\,\ln\left(\rho(\mu)\right)\right],
\end{equation}
from which we can read off the entropy corresponding to the density field
$\rho$. 

Putting this all together we find
\begin{equation}
Z_N(\zeta) = A_N\int d[\rho]  \delta\left(\int d\mu \ \rho(\mu) -1\right)
\exp\left[-N\int 
d\mu \ \rho(\mu)\ln\left(\rho(\mu)\right) - {\beta N^2} {\cal E}[\rho]\right],
\label{func2} 
\end{equation}
where $A_N$ is a prefactor and the delta function enforces the
normalization condition in Eq.  (\ref{normf}). One simple way to
incorporate this delta function constraint is to introduce a Lagrange
multiplier $C$ (corresponding to writing the delta function constraint
in the Fourier representation) and rewrite the partition function as
\begin{equation}
Z_N(\zeta)= A'_N \int dC\, d[\rho] \exp\left[-\beta N^2 \Sigma[\rho]- N(1-{\beta\over 2})\int
d\mu \ \rho(\mu)\ln\left(\rho(\mu)\right)\right],
\label{func3}
\end{equation}
where $A'_N$ is a constant and the `renormalized' action
is explicitly
\begin{equation}
\Sigma[\rho]= \frac{1}{2}\,\int d\mu\, \mu^2\, \rho(\mu) -
\frac{1}{2}\,\int d\mu\, d\mu' \rho(\mu)\, \rho(\mu')\,
\ln(|\mu-\mu'|) + C \left[\int d\mu \rho[\mu] -1\right].
\label{action0}
\end{equation}
The $O(N)$ term in Eq. (\ref{func3}) includes both the entropy of the density
field and the self energy subtraction following the Dyson
prescription discussed before. While this prescription is difficult to justify rigorously,
we need not pursue this issue here since we are interested only in the leading $O(N^2)$
behavior.

The functional integral in Eq. (\ref{func3}) can thus be
evaluated by the saddle point method in the variable $N^2$ and the
term of order $N$ coming from the entropy and self energy
subtraction is negligible in the large
$N$ limit. We thus find that in the saddle point analysis, for large
$N$,
\begin{equation}
Z_N(\zeta) = \exp\left[-{\beta N^2} \Sigma[\rho_c] + O(N)\right]
\label{pf} 
\end{equation}
where $\rho_c(\mu)$ is the density field that minimizes the action $\Sigma[\rho]$ in
Eq. (\ref{action0}). Minimizing  $\Sigma[\rho]$, it follows
that $\rho_c(\mu)$ satisfies the integral
equation
\begin{equation}
\frac{\mu^2}{2} + C = 
\int_z^\infty d\mu'\ \rho_c(\mu') \ln\left(|\mu-\mu'|\right),
\label{eqvar}
\end{equation}
where we have introduced the scaled variable $z = \zeta/\sqrt{N}$.
The normalization and 
boundary conditions for $\rho_c(\mu)$, in terms of the scaled variable $z$, are  
\begin{equation}
\int_z^\infty d\mu\,  \rho_c(\mu) = 1 ;\ \ \ \rho_c(\mu) = 0\ {\rm for}\  \mu < z.
\label{eqnf}
\end{equation}

The partition function then reads
\begin{equation}
Z_N(z) = \exp\left(-\beta\, N^2\, S(z) + O(N)\right).\label{eqst}
\end{equation}
where
\begin{equation}
S(z) = \min_\rho\left\{ \Sigma[\rho]\right\} = \Sigma[\rho_c]
\label{action1}
\end{equation}

Let us remark here that in the unconstrained case ($z=-\infty$) the 
Wigner semi-circle law can also be obtained from the saddle-point method \cite{mehta}.
However to our knowledge the constrained problem has never been analyzed
using this approach. In addition, the effective free energy is not extensive: 
it scales as $N^2$ rather than $N$ due to the long-range nature of the 
logarithmic inter-particle interaction. The entropy term
is extensive and is  contained in the $O(N)$ term in Eq. (\ref{eqst}).
However the entropy is subdominant, and the free energy is  dominated by the energetic 
component.

Differentiating 
Eq. (\ref{eqvar}) with respect to $\mu$ we get
\begin{equation}
\mu = {\cal P}\int_z^\infty d\mu'\ \frac{\rho_c(\mu')}{\mu -\mu'},
\label{eqdvar}
\end{equation}
where $\cal{P}$ indicates the Cauchy principle part. It is convenient to
shift the variable $\mu$ by writing
\begin{equation}
\mu = z + x
\end{equation}
and where $x\ge 0$ is now positive and $x=0$ denotes the location
of the infinite barrier. In terms of the shifted variable, we denote
the density field as
\begin{equation}
\rho_c(\mu=x+z)= f(x;z).
\label{shiftden}
\end{equation}
Eq. (\ref{eqdvar}) then becomes 
\begin{equation}
x + z = {\cal P}\int_0^\infty dx'\ \frac{ f(x';z)} {x -x'}.
\label{inteq1}
\end{equation}
This integral equation is of the general form
\begin{equation}
g(x)= (H_+ f)(x)= {\cal P}\int_0^\infty dx'\ \frac{f(x')}{x -x'},
\label{hilbert1}
\end{equation}
with $g(x)=x+z$. 
The right hand side (rhs) of the above equation is just the
semi-infinite Hilbert transform of the function $f(x)$.
The main technical challenge is to invert this half Hilbert transform.
Fortunately this inversion can be done for arbitrary $g(x)$ using Tricomi's 
theorem~\cite{hhilbert}.
To apply Tricomi's theorem, we first assume, to be verified aposteriori,
that $f(x)$ has a finite support over $[0,L]$, so that the
integral in Eq. (\ref{inteq1}) can be cut-off at $L$ at the
upper edge. Then the solution for $f(x)$ is given by Tricomi's theorem~\cite{hhilbert} 
\begin{equation}
f(x) = -\frac{1}{\pi^2\sqrt{x(L-x)}}\left\{
{\cal P}\int_0^L dx'\ \sqrt{x'(L-x')}\,\frac{g(x')}{x-x'} + C'\ \ \right\}
\label{tricomi1}
\end{equation}
where $C'$ is an arbitrary constant which can be fixed by
demanding that $f(x)$ vanishes at $x=L$. Applying this formula to our 
case with $g(x)=x+z$ and performing the integral in Eq. (\ref{tricomi1}) using
Mathematica, we get 
\begin{eqnarray}
f(x;z) &=& \frac{1}{2\pi \sqrt{x}}\,\sqrt{L(z)-x}\, \left[L(z)+2x + 2z\right],
\quad x\in[0,L(z)]\nonumber \\
f(x;z) &=& 0,\quad \ x<0 \ {\rm and } \ x > L(z) 
\label{eqfz}
\end{eqnarray}
where we have made the $z$ dependence of $L(z)$ explicit. The value
of $L(z)$ is now determined from the normalization of $f(x;z)$, i.e.,
$\int_0^{L(z)} f(x;z) dx=1$ and one gets 
\begin{equation}
L(z) = \frac{2}{3}\,\left[ \sqrt{z^2 + 6} -z\right]
\label{lz}
\end{equation}
which is always positive.
We also see that in terms of the variable $\mu=x+z$,
the support of $\rho_c(\mu)$ is between the barrier location at $z$ and  
an upper edge at $\mu=z+L(z)=\frac{2}{3}\, \left[ \sqrt{z^2 + 6} + 2z\right]$. 

Let us make a few observations about the constrained charge density 
$\rho_c(\mu=x+z)=f(x;z)$ in Eqs. (\ref{eqfz}) and (\ref{lz}).
\vskip 0.2cm

$\bullet$ Physically, the charge density $f(x;z)$ must be positive for all $x$ including
$x=0$. Now, as $x\to 0$, i.e., as one approaches the infinite barrier
from the right, it follows from Eq. (\ref{eqfz}) that $f(x;z)$ diverges
as $x^{-1/2}$, i.e., the charges accumulate near the barrier. However, for it to remain 
positive at $x=0$, it follows from Eq. (\ref{eqfz}) that we must have 
$L(z)+2z\ge 0$. This happens, using the expression of $L(z)$ from Eq. (\ref{lz}),
only for $z\ge -\sqrt{2}$. Thus the result in Eq. (\ref{eqfz}) 
is valid 
only for $z\ge -\sqrt{2}$. To understand the significance of $z\ge -\sqrt{2}$, 
we note that when $z=-\sqrt{2}$, i.e., when the barrier is placed exactly at the leftmost
edge of the Wigner sea, we recover the Wigner semi-circle law from Eqs. (\ref{eqfz})
and (\ref{lz}).
We get $L(z=-\sqrt{2})= 2\sqrt{2}$ (the support of the full semi-circle), i.e.,
$f(x;-\sqrt{2})$ is nonzero for $0\le x\le 2\sqrt{2}$. In terms of the original
variable $\mu=x+z=x-\sqrt{2}$, $\rho_c(\mu)$ is nonzero in the region 
$-\sqrt{2}\le \mu\le \sqrt{2}$ and is given by,
within this sea to the right of the barrier, in terms of the original variable $\mu$
\begin{equation}
\rho_c(\mu)= \frac{1}{\pi} \sqrt{2-\mu^2}. 
\label{wig2}
\end{equation}
When $z$ becomes smaller than $-\sqrt{2}$, the solution in Eq. (\ref{eqfz})
remains unchanged from its semi-circular form at $z=-\sqrt{2}$. In other
words, for $z\le -\sqrt{2}$, the solution sticks to its form at $z=-\sqrt{2}$. 
Physically, this means that if the barrier is placed to the left of the 
lower edge of the Wigner sea, it has no effect on the charge distribution.

\vskip 0.2cm
$\bullet$ Note that the above fact, in conjunction with Eq. (\ref{action1}),
indicates that 
\begin{equation}
S(z)= S(-\sqrt{2}) \quad {\rm for}\,\, {\rm all} \,\, z\le -\sqrt{2}
\label{action2}
\end{equation}
including, in particular
\begin{equation}
S(-\infty)= S(-\sqrt{2})
\label{action3}
\end{equation}
a result that we will use later.

\begin{figure}
\includegraphics[width=.45\textwidth]{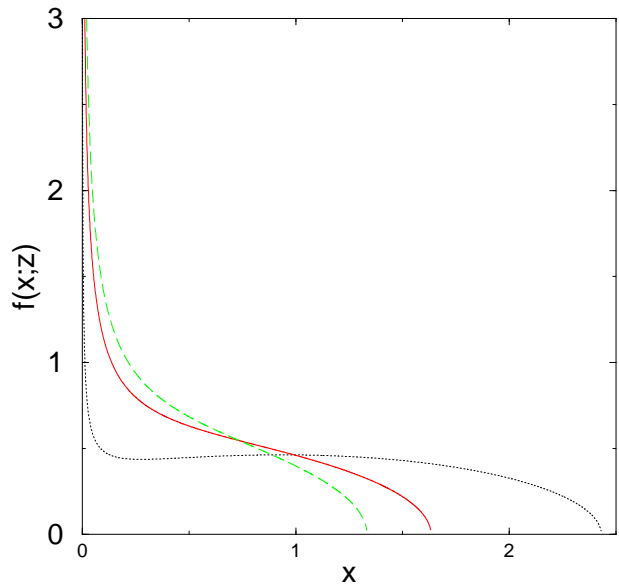}
\caption{Plots of the density of states $f(x;z)$ as a function of the 
shifted variable $x$ for $z=-1$ (dotted), $z= 0$ (solid), and $z=0.5$ 
(dashed).  }
\label{fplot}
\end{figure}

\vskip 0.2cm
$\bullet$ The charge density $\rho_c(\mu=x+z)=f(x;z)$ changes its shape in an interesting 
manner 
as one changes the barrier location $z$. 
For $z> -\sqrt{2}$ the global maximum of $f(x;z)$ is always at $x=0$ 
(at the barrier) where it has a $1/\sqrt{x}$ integrable
singularity. For $-\sqrt{2}< z < -\sqrt{3/4}$, the density is
non-monotonic and in addition to the square root divergence at $x=0$, $f(x;z)$
develops a local minimum and a local maximum respectively at
$x= (L\mp\sqrt{-3L^2-8zL})/4$. For $z>-\sqrt{3/4}$, the density
$f(x;z)$ decreases monotonically with increasing $x$. 
In Fig. (\ref{fplot}) we show 
the behavior of $f(x;z)$  for three representative values
of $z$. An interesting point about the eigenvalue distribution $f(x;z)$ given 
Eq. (\ref{eqfz}) is that, for $z> -\sqrt{2}$ there is a 
strong accumulation of eigenvalues at the barrier location $x=0$ or equivalently
at $\mu=z$. In the case $z=0$ if  one thinks of the eigenvalues as being 
associated about the  vacuum of a (stable) field theory then there is an 
accumulation of modes of mass close to zero, a fact that may have 
consequences in the context of anthropic principal based string theory
or in other physical systems where only stable configurations can be observed.  

\subsection{Large Deviations of the Maximum or the Minimum}

Having computed the constrained charged density $\rho_c(\mu)$, we next 
calculate the action $S(z)=\Sigma[\rho_c]$ at the saddle point.
In order to calculate $\Sigma[\rho_c]$ from Eq. (\ref{action0}), we 
will use the explicit saddle point solution $\rho_c(\mu=x+z)= f(x;z)$
obtained in Eq. (\ref{eqfz}). In fact, one can simplify the expression
of the saddle point action by using the integral equation (\ref{eqvar}) satisfied
by the saddle point solution. We multiply Eq. (\ref{eqvar}) by $\rho_c(\mu)$
and integrate over $\mu$. Using the normalization $\int \rho_c(\mu)\, d\mu=1$
one gets
\begin{equation}
\int d\mu\, d\mu'
\rho_c(\mu)\, \rho_c(\mu')\, \ln(|\mu-\mu'|) = C + \frac{1}{2}\, \int \mu^2\, \rho_c(\mu)\, 
d\mu
\label{manip1}
\end{equation}
where $C$ is the Lagrange multiplier to be determined.
Substituting this result in Eq. (\ref{action0}) and the fact, $\int \rho_c(\mu)\, d\mu=1$
we get
\begin{equation}
\Sigma[\rho_c]= -\frac{1}{2}\, C + \frac{1}{4}\, \int \mu^2\, \rho_c(\mu)\, d\mu.
\label{manip2}
\end{equation}
To determine the Lagrange multiplier $C$, we put
$\mu=z$ in Eq. (\ref{eqvar}).
In the integral on the rhs of Eq. (\ref{eqvar}) we then make the usual shift, $\mu'=z+x'$
and use $\rho_c(\mu')= f(x';z)$ to get
\begin{equation}
C = -\frac{1}{2}\, z^2 +  \int_0^{L(z)} dx'\ f(x';z) \ln\left(x'\right).
\label{lagrange1}
\end{equation}
where $f(x;z)$ is explicitly given in Eq. (\ref{eqfz}) and $L(z)$ in Eq. (\ref{lz}). 
Substituting the expression of $C$
in Eq. (\ref{manip2}) and $\rho_c(\mu')= f(x';z)$ we finally get
\begin{equation}
\Sigma[\rho_c]= \frac{1}{4}\, z^2-\frac{1}{2}\,\int_0^{L(z)} dx\, \ln(x)\, f(x;z)+ 
\frac{1}{4}\,\int_0^{L(z)} dx\, (x+z)^2\, f(x;z)
\label{manip3}
\end{equation}
with $f(x;z)$ given explicitly in Eq. (\ref{eqfz}). The integrals can again be
performed explicitly using Mathematica and using 
Eq. (\ref{action1})
we get the following expression for the
saddle point action 
\begin{equation}
S(z) = \frac{1}{216}\left[ 72 z^2 -2z^4 +(30 z + 2z^3)
\sqrt{6 +z^2}+ 27\left( 3 + \ln(1296) - 4
\ln\left(-z + \sqrt{6 +z^2}\right)\right)\right].
\label{saddle1}
\end{equation} 

The partition function $Z_N(\zeta=\sqrt{N} z)$ is then given by Eq. (\ref{eqst}). 
Note also, using Eq. (\ref{action3}), we have 
\begin{equation} 
Z_N(-\infty)= \exp\left(-\beta\, N^2\, S(-\sqrt{2}) + O(N)\right), 
\label{action4} 
\end{equation} 
where $S(-\sqrt{2})= (3+\ln(4))/8$ from Eq. (\ref{saddle1}).
Taking 
the ratio in Eq. (\ref{qnz3}) gives us $Q_N(\zeta)$, the probability $Q_N(\zeta)$ 
that all eigenvalues are to the right of $\zeta = \sqrt{N} z$ 
\begin{equation} 
Q_N(\zeta) = \exp\left[-\beta\, N^2\, \theta\left(\frac{\zeta}{\sqrt{N}}\right) +O(N)\right]
\label{qnz5} 
\end{equation} 
where  $ \theta(z) = S(z)-S(-\sqrt{2})$ 
is given by
\begin{equation} 
\theta(z)= \frac{1}{108} 
\left[36 z^2 -z^4 + (15 z + z^3)\sqrt{6 + z^2}  
+ 27 
\left(\ln(18) -2\ln\left(-z + \sqrt{6 +z^2}\right)\right)\right]. 
\label{eqtheta} 
\end{equation}

The probability that all eigenvalues are positive (or negative) is simply
\begin{equation}
P_N= Q_N(\zeta=0)\approx \exp[-\beta\, \theta(0)\,N^2]
\label{pnasymp}
\end{equation}
where   
\begin{equation} 
\theta(0) = \frac{\ln(3)}{4}=0.274653.... 
\label{theta01}
\end{equation}

Finally let us turn to the large deviation function 
associated with
large negative fluctuations of $\sim -O(N^{1/2})$ of $\lambda_{\rm max}$ to the left of 
its
mean value $\sqrt{2\,N}$.
Substituting the expression for $Q_N(\zeta)$ from Eq. (\ref{qnz5})
in Eq. (\ref{maxmin1}) we get
\begin{equation}
{\rm Prob}[\lambda_{\rm max}\le t, N]= Q_N(\zeta=-t)=\exp\left[-\beta N^2 
\theta\left(-\frac{t}{\sqrt{N}}\right) +O(N)\right].
\label{minmax2}
\end{equation}
Noting that $\langle \lambda_{\rm max}\rangle=\sqrt{2N}$, it is useful
to center the distribution around the mean and rewrite Eq. (\ref{minmax2}) as
\begin{equation}
{\rm Prob}[\lambda_{\rm max}\le t, N]= \exp\left[-\beta N^2 \Phi\left( 
\frac{\sqrt{2N}-t}{\sqrt{N}} \right) \right],
\label{ldf1}
\end{equation}
where the large deviation function $\Phi(y)= \theta(y-\sqrt{2})$ 
with $\theta(z)$ given explicitly in Eq. (\ref{eqtheta}).

One can easily work out the asymptotic behavior of $\Phi(y)$ for small
and large $y$. For example, it is easy to see that
\begin{eqnarray}
\Phi(y) &\approx & \frac{y^3}{6\sqrt{2}}\quad\, {\rm as}\quad y\to 0 \nonumber \\
        &\approx & \frac{y^2}{2}\quad\, {\rm as}\quad y\to \infty
\label{ldfasymp}
\end{eqnarray}
In particular, when $\sqrt{2N}-t << \sqrt{N}$, i.e, we are rather close to the
right edge $\langle \lambda_{\rm max} \rangle=\sqrt{2N}$ 
of the Wigner sea 
(see Fig. 1), it follows that the scaling variable $y= (\sqrt{2N}-t)/\sqrt{N} << 1$.
Hence substituting the small $y$ behavior of $\Phi(y)\approx y^3/{6\sqrt{2}}$ from
Eq. (\ref{ldfasymp}) in Eq. (\ref{ldf1}), it follows that in this regime
\begin{equation}
{\rm Prob}[\lambda_{\rm max}\le t, N] \approx \exp\left[-\frac{\beta}{24} 
\left(|\sqrt{2} 
N^{1/6} (t-\sqrt{2N})|\right)^3\right].
\label{ldfasymp1}
\end{equation}
Note that this matches exactly with the left tail behavior of the
Tracy-Widom limiting distribution for all the three cases $\beta=1$,
$2$ and $4$ \cite{comment1}. For example, for $\beta=2$, one can
easily verify by comparing Eqs. (\ref{ldfasymp1}) and
(\ref{asymp1}). This is to be expected because the Tracy-Widom
distribution describes the distribution of $\lambda_{\rm max}$ around
its mean $\sqrt{2N}$ over a scale $\sim O(N^{-1/6})$. If we want to
investigate the probability of negative fluctuations of order
$(\sqrt{2N}-t)>> N^{-1/6}$, we need to look at the left tail of the
Tracy-Widom distribution. On the other hand, those negative
fluctuations of order $N^{-1/6}<< (\sqrt{2N}-t)<< \sqrt{N}$ are
described by the small argument behavior of the large deviation
function.  These two behaviors thus should smoothly match. As we
verified above, they do indeed match smoothly, thus providing another
confirmation of our exact result.  Moreover, our large deviation
function provides an alternative way to compute the left tail of the
Tracy-Widom distribution for all $\beta$.

\section{Coulomb Gas Bounded by two walls: the joint probability distribution of $\lambda_{\rm 
min}$ and $\lambda_{\rm max}$}

In this section we compute, by the Coulomb gas method, the probability
$R_N(\zeta_1,\zeta_2)$ that all eigenvalues are in the interval
$[\zeta_1,\zeta_2]$ where $\zeta_2\ge \zeta_1$. Evidently, in the
limit $\zeta_2\to \infty$, $R_N(\zeta,\infty)= Q_N(\zeta)$ which was
computed in the previous section.  Clearly, $R_N(\zeta_1,\zeta_2)$ is
also the cumulative probability that $\lambda_{\rm min}\ge \zeta_1$
and $\lambda_{\rm max}\le \zeta_2$, i.e.,
\begin{equation}
R_N(\zeta_1,\zeta_2)= {\rm Prob}\left[ \zeta_1\le \lambda_1\le \zeta_2,\,
\zeta_1\le \lambda_2\le \zeta_2,\ldots, \zeta_1\le \lambda_N\le \zeta_2\right]=
{\rm Prob}\left[\lambda_{\rm min}\ge \zeta_1,\, \lambda_{\rm max}\le \zeta_2\right].
\label{rn1}
\end{equation}
In other words, $R_N(\zeta_1,\zeta_2)$ provides the joint probability distribution
of the minimum and the maximum eigenvalue. By definition,
\begin{equation}
R_N(\zeta_1,\zeta_2)=
\int_{\zeta_1}^{\zeta_2}\ldots\int_{\zeta_1}^{\zeta_2}
P(\lambda_1,\lambda_2,\ldots, \lambda_N)\,d\lambda_1\,d\lambda_2\ldots
d\lambda_N
\label{rn2}
\end{equation}
where $P$ is the joint pdf in Eq. (\ref{pdf}).
As in the previous section, this multiple integral can be written as a ratio
of two partition functions
\begin{equation}
R_N(\zeta_1,\zeta_2)= \frac{\Omega_N(\zeta_1,\zeta_2)}{Z_N(-\infty)}
\label{rn3}
\end{equation}
where 
\begin{equation}
\Omega_N(\zeta_1,\zeta_2)= \int_{\zeta_1}^{\zeta_2}\ldots\int_{\zeta_1}^{\zeta_2}
\, \exp\left[-\frac{\beta}{2}\left(\sum_{i=1}^N\lambda_i^2
-\sum_{i\ne j}\ln(|\lambda_i-\lambda_j|)\right)\right]\,d\lambda_1\,d\lambda_2\ldots 
d\lambda_N.
\label{dwpf1}
\end{equation}
and $Z_N(-\infty)=\Omega_N(-\infty,\infty)$ is the same normalization
constant as in Eq.  (\ref{qnz3}).  Thus, $\Omega_N(\zeta_1,\zeta_2)$
represents the partition function of the Coulomb gas that is
sandwiched in the region $[\zeta_1,\zeta_2]$ bounded by the two hard
walls at its boundaries.

We then evaluate this partition function in the large $N$ limit using
the saddle point method.  The formalism is exactly same as in the
previous section. We first define a counting function $\rho(\mu)$ that
is nonzero only in the region $\frac{\zeta_1}{\sqrt{N}}\le \mu\le
\frac{\zeta_2}{\sqrt{N}}$ and is zero outside. The rest of the
calculation is similar as in the previous section, except that all the
integrals run over the region $\mu\in [z_1,z_2]$ where
$z_1=\zeta_1/\sqrt{N}$ and $z_2=\zeta_2/\sqrt{N}$. The action
$\Sigma[\rho]$ is exactly as in Eq. (\ref{action0}).  Thus the
partition function, in the large $N$ limit, behaves as
\begin{equation}
\Omega_N(\zeta_1,\zeta_2)= \exp\left[-\beta\, N^2\, \Sigma[\rho_c] +O(N)\right]
\label{dwpf2}
\end{equation}
where the saddle point density $\rho_c(\mu)$, 
in terms of scaled 
variables
$z_1= \zeta_1/\sqrt{N}$ and $z_2=\zeta_2/\sqrt{N}$, satisfies the integral equation
\begin{equation}
\mu = {\cal P}\int_{z_1}^{z_2} d\mu'\ \frac{\rho_c(\mu')}{\mu -\mu'},
\label{dweqdvar}
\end{equation}
where $\cal{P}$ indicates the Cauchy principle part.
Next we introduce the shift
\begin{equation}
\mu= z_1 +x
\end{equation}
and define $W=z_2-z_1$. Since $z_1\le \mu\le z_2$, it follows that
$0\le x\le W$.  Thus $x=0$ denotes the 
location of the left barrier
and $x=W$ denotes the location of the right barrier. In terms of the shifted variable,
we rewrite the density field as
\begin{equation}
\rho_c(\mu=z_1+x)= f(x; z_1, W).
\label{dwshiftden}
\end{equation}  
Eq. (\ref{dweqdvar}) then reduces to the integral equation
\begin{equation}
x + z_1 = {\cal P}\int_0^W dx'\ \frac{ f(x';z_1,W)} {x -x'}.
\label{dwinteq1}
\end{equation}

The integral equation (\ref{dwinteq1}) can again be solved using
Tricomi's theorem.  Note that this equation has almost similar form as
Eq. (\ref{inteq1}) except that the integral on the rhs of
Eq. (\ref{dwinteq1}) runs up to $W$. From the solution of
Eq. (\ref{inteq1}) presented in Eq. (\ref{eqfz}) we learned that the
density $f$ is nonzero only for $0\le x\le L(z)$ and is zero for
$x>L(z)$ where $L(z)= \frac{2}{3}\,\left[ \sqrt{z^2 + 6}
-z\right]$. So, comparing to Eq. (\ref{dwinteq1}) we see that there
are two possibilities:
\vskip 0.2cm

\noindent (i)  If 
$W= z_2-z_1 > L(z_1)$, the solution $f(x;z_1, W)$ will be exactly the same
as $f(x;z_1)$ presented in Eq. (\ref{eqfz}). In this case, the Coulomb
gas does not feel the presence of the right barrier at $z_2$. In other words,
the solution $f(x;z_1, W)=f(x;z_1, \infty)$ is completely independent of  
$W$ and one can effectively put $W\to \infty$, i.e., put the right barrier at
infinity. Thus in the case, the charge density diverges as $x^{-1/2}$ at
the left barrier and vanishes at $x=L(z_1)$.
\vskip 0.2cm

\noindent (ii) If $W= z_2-z_1< L(z_1)$, then the solution will be given by
Eq. (\ref{tricomi1}) with $L$ replaced by $W$. Using $g(x)=x+z_1$ in 
Eq. (\ref{tricomi1}) and performing the integral on the rhs we get
\begin{equation}
f(x;z_1,W)= \frac{1}{8\pi \sqrt{x(W-x)}}\left[W^2+4W(x+z_1)-8x(x+z_1)+B'\right]
\label{dwtricomi1}
\end{equation}
where $B'$ is an arbitrary constant. The normalization condition, $\int_0^{W} 
f(x;z_1,W)\,dx=1$, fixes the constant $B'=8$. In this case, the charge density
diverges (with a square root singularity) at the locations of both the left
barrier ($x=0$) and the right barrier ($x=W$).

Thus, putting (i) and (ii) together, we find that the solution for the
equilibrium charge density $f(x;z_1,W)$ for the Coulomb gas sandwiched
between two barriers is given by  
\begin{equation}
f(x;z_1,W)= \frac{1}{8\pi \sqrt{x(l-x)}}\left[l^2+4l(x+z_1)-8x(x+z_1)+8\right],\quad {\rm 
for}\quad 0\le x\le l
\label{dwsol1}
\end{equation}
where 
\begin{equation}
l= {\rm min}\left[W=z_2-z_1,\, L(z_1)= \frac{2}{3}\,\left(\sqrt{z_1^2+6}-z_1\right)\right].
\label{dwl1}
\end{equation} 
Clearly, in the limit $W\to \infty$, we indeed recover the results of
the previous section.  Summarizing, if one fixes the left barrier at
$z_1$ and varies the position of the right barrier $z_2$ (equivalently
by varying the distance $W=z_2-z_1$ between the two walls), one finds
that the charge density at the left barrier always diverges. On the
other hand, the behavior of the density near the right barrier
undergoes a sudden change as $W$ increases beyond a critical value
$W_c=L(z_1)= \frac{2}{3}\,\left(\sqrt{z_1^2+6}-z_1\right)$. The
density at the right wall diverges as long as $W<W_c$, i.e., $z_2<
z_1+ L(z_1)$. But when $W>W_c$ or equivalently $z_2>z_1+L(z_1)$, the
charge density goes to zero at the right edge of the support at
$L(z_1)<W$.

\begin{figure}
\includegraphics[width=.45\textwidth]{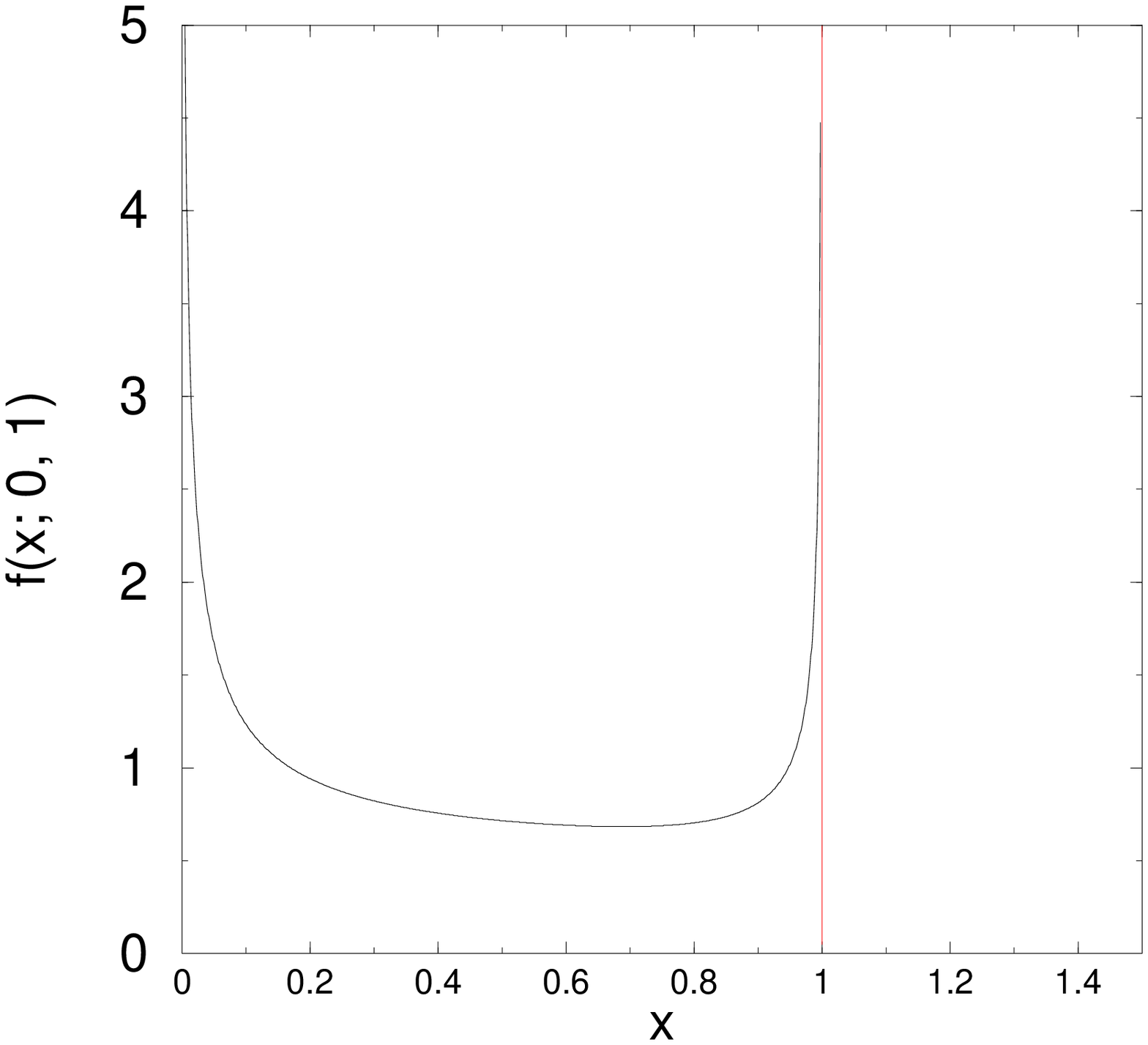}\hfill
\includegraphics[width=.45\textwidth]{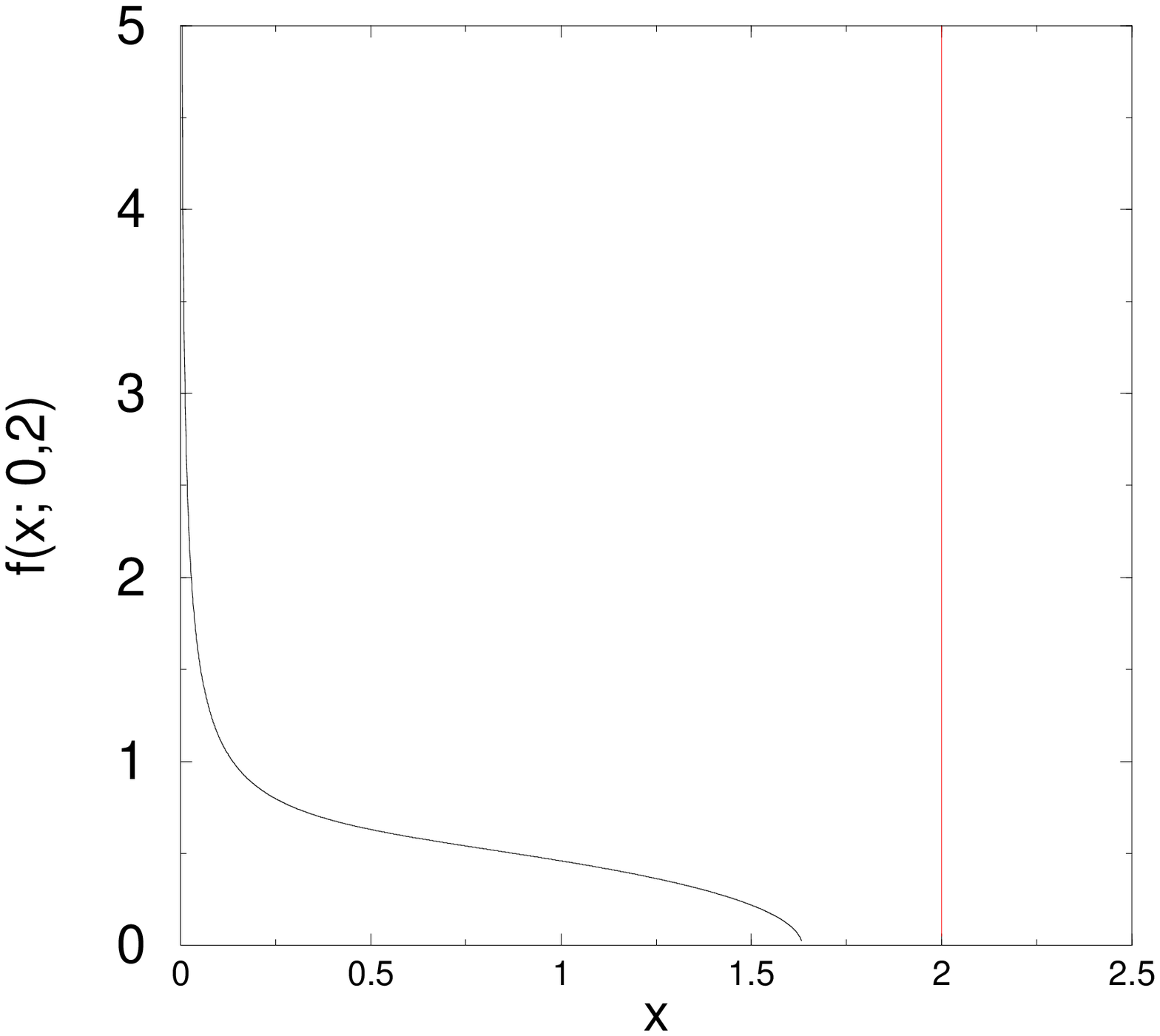}
\caption{The charge density $f(x;z_1,W)$ for $z_1=0$ and $W=1$ and
$W=2$ respectively.  When the left wall location $z_1=0$,
$L(0)=\sqrt{8/3}=1.63299$. Thus the critical value of the distance $W$
between the walls is $W_c=1.63299$. On the left panel, $W=1<W_c$
(subcritical) where the charge density diverges (square root
singularity) at the location of second wall $x=W$. On the right panel,
$W=2>W_c$ (supercritical) where the charge density goes to zero as
$x\to W_c=L(0)< W$.}
\end{figure}

%\begin{figure}
%\subfigure{\includegraphics[width=.45\textwidth]{dwden1.eps}}
%\subfigure{\includegraphics[width=.45\textwidth]{dwden2.eps}}
%\end{figure}

Having determined the charge density $\rho_c(\mu)= f(x=\mu+z_1; z_1,W)$, the 
saddle point action $\Sigma[\rho_c]$, is then determined via the following equation
that is analogous to Eq. (\ref{manip3})
\begin{equation}
\Sigma[\rho_c]= \frac{1}{4}\, z_1^2-\frac{1}{2}\,\int_0^{l} dx\, \ln(x)\, f(x;z_1,W)+
\frac{1}{4}\,\int_0^{l} dx\, (x+z_1)^2\, f(x;z_1,W)
\label{dwmanip}
\end{equation}
where $W=z_2-z_1$ and $f(x;z_1,W)$ and $l$ are given respectively in
Eqs. (\ref{dwsol1}) and (\ref{dwl1}).  Denoting the saddle point
action $S(z_1,W)= \Sigma[\rho_c]$ and evaluating explicitly the
integrals in Eq. (\ref{dwmanip}) we get for $W< L(z_1)$
\begin{equation}
S(z_1,W)= \frac{1}{32}\,\left[32\, \ln(2)-16\, \ln(W)+16\, z_1^2
+6\, W^2 + 16\, W z_1 -2\, W^2z_1^2 -2\, W^3 z_1-\frac{9}{16}\, W^4\right].
\label{dwac1}
\end{equation}
For $W>L(z_1)$, $S(z_1,W)$ becomes independent of $W$ and sticks to
its value $S(z_1, L(z_1))$. On the other hand, we know that when $W\to
\infty$, $S(z_1,\infty)$ must be equal to the action $S(z_1)$ for a
single wall as given in Eq. (\ref{saddle1}).  Indeed, one can check
explicitly that $S(z_1, L(z_1))=S(z_1)$, thus confirming the
expectation.

The partition function, in terms of the scaled variables $z_1$
and $z_2$, then 
follows from Eq. (\ref{dwpf2})
\begin{equation}
\Omega_N(z_1,z_2)= \exp\left[-\beta\, N^2\, S(z_1,W)+O(N)\right].
\label{dwpf3}
\end{equation} 
Note that the denominator $Z_N(-\infty)$ in Eq. (\ref{rn2}) is still
given by Eq. (\ref{action4}) where $S(-\sqrt{2})= (3+\ln(4))/8$. Hence
taking the ratio in Eq.  (\ref{rn2}) and using Eq. (\ref{dwpf3}) we
get the joint probability $R_N(\zeta_1,\zeta_2)$ for large $N$
\begin{equation}
R_N(\zeta_1,\zeta_2)= \exp\left[-\beta\, N^2\, 
\Psi\left(\frac{\zeta_1}{\sqrt{N}},\frac{\zeta_2}{\sqrt{N}}\right)+ O(N)\right],
\label{psi1}
\end{equation}
where 
\begin{equation}
\Psi(z_1,z_2)= S(z_1,W=z_2-z_1)- \frac{3+\ln(4)}{8}
\label{psi2}
\end{equation}
with $S(z_1,W)$ given by Eq. (\ref{dwac1}). One can check easily that
when the second wall moves to infinity, i.e., $z_2\to \infty$, 
$\Psi(z_1,\infty)=\theta(z_1)$
where $\theta(z)$ is given in Eq. (\ref{eqtheta}). Thus, Eq. (\ref{psi1}) for the
joint distribution of $\lambda_{\rm min}$ and $\lambda_{\rm max}$ is a generalization
of Eq. (\ref{qnz5}) that describes only the distribution of $\lambda_{\rm min}$.

\section{Numerical Results}

The reader will realize that the numerical confirmation of the 
analytical results of the previous section is a delicate and potentially 
computation intensive task. For simplicity we will restrict our selves to
the case of the GOE but the methods used can be extended to the 
other ensembles. The simplest way to compute the probability that
all eigenvalues are greater than some value is to numerically generate 
matrices from the required ensemble, diagonalize them and then count the 
number $m_+$ that satisfy the eigenvalue constraint required. 
However because of the 
order $N^2$ suppression of this probability found here, for large 
$N$ the number of matrices $m$ that one would need to generate before seeing 
a single matrix satisfying the constraint is huge. The estimate
for the probability that all eigenvalues are positive in this method is  
given by
\begin{equation}
Q_N(0) = \frac{m_+}{m}.
\end{equation}     
In \cite{AE} an approximate argument for the GOE
($\beta =1$) was made yielding $\theta = 1/4$ and a subsequent numerical study 
on matrices  up to $7\times 7$ with an $N$-dependent fit $\theta= aN^\alpha$ 
yielded $\alpha = 2.00387$ and $a= 0.3291$ \cite{comment2} was found. 
However given that there are  $O(N)$ corrections and the size of the systems
studied are so small this fit cannot be taken too seriously. In fact one can  
use the Coulomb gas representation of the eigenvalues of Gaussian ensembles
in order to numerically compute $\theta(0)$ for much larger values of $N$. 
However as a test of this method for smaller values of $N$ we may adopt the 
direct enumeration approach of \cite{AE} but slightly improve it to gain a 
few extra values of $N$. 

If the all the eigenvalues of a matrix  are positive then for any vector
${\bf v}$ we must have that 
\begin{equation}
({\bf v}, M{\bf v}) > 0 \label{sprod}
\end{equation}
In particular if we choose the vector ${\bf v}$ to be one of the $N$ basis 
vectors ${\bf e}_i$ then Eq. ({\ref{sprod}) implies that 
\begin{equation}
({\bf e}_i, M{\bf e}_i) = M_{ii} > 0,
\end{equation}
and so if $M$ is positive (in the operator sense), all of the diagonal 
elements must be positive. The estimation of $Q_N(0)$ can thus be slightly
improved by increasing the chances of seeing a positive matrix by 
forcing the diagonal elements to be positive. With respect the the 
simplest form of enumeration, the matrices $M^*$ generated are the same
as those for the GOE but the diagonal elements are replaced with their
absolute value. The probability that a given matrix $M$ has all diagonal
elements positive is $1/2^N$, thus if $m^*_+$ denotes the number of 
these so generated matrices (with positive diagonal elements) then  the 
estimation of the probability that a GOE matrix is positive is given by:
\begin{equation}
Q_N(0) = \frac{1}{2^N} \frac{m^*}{m}.
\end{equation} 
For small $N$ this method thus appreciably increases the probability
of generating positive matrices and thus enhances the accuracy of the
estimate for $Q_N(0)$.  Even so using this method it is virtually
impossible to obtain meaningful results for $N>8$. The results
obtained by this modified enumeration method are shown on
Fig. (\ref{qplot}) (squares).

For large $N$  it is in fact much better to evaluate
$Q_N(0)$ directly from Eq. (\ref{qnz3}) via a Monte Carlo method. We note that
we can write 
\begin{equation}
\frac{1}{(2\pi)^{\frac{N}{2}}} Z(-\infty) = \langle G(\lambda)\rangle,
\end{equation}
where $G(\lambda) = \prod_{i<j} |\lambda_i-\lambda_j|^\beta$ and the angled
bracket indicates the average is over $\lambda_i$ taken to be
independent and Gaussian of zero mean and unit variance. The term
$Z(0)$ can also be related to the expectation over $\lambda_i$ which
are similarly independent and Gaussian of zero mean and unit variance
but conditioned to be positive, and we denote the average of with
respect to these variables by $\langle\;\cdot\;\rangle_+$. We find that
\begin{equation}
\frac{2^N}{(2\pi)^{\frac{N}{2}}}{ Z(0)} = \langle G(\lambda)\rangle_+,
\end{equation}
the left-hand side now has has the form of a conditional average, the factor
of $2^N$ giving the correct normalization for the conditioned probability 
distribution. Putting all this
together yields
\begin{equation}
Q_N(0) =\frac{1}{2^N} \frac{ \langle G(\lambda)\rangle_+}{\langle G(\lambda)\rangle}
\end{equation}
The two expectation values can be computed via Monte Carlo sampling,
the unconditioned one by using Gaussian random variables and the
second trivially by using the absolute value of Gaussian random
variables. Shown in Fig. (\ref{qplot}) is a plot of $\ln(Q_N(0))$ for the
GOE ($\beta = 1$) ensemble
measured as described above (circles), we see that the agreement for
small $N$ with the results obtained by modified enumeration approach
is excellent. For larger values of $N$ we have used $5\times 10^8$
Monte Carlo samplings and there is a significant amount of fluctuation
as indicated by the error bars.  The Monte Carlo results were fitted
using the fit a fit $ax^2 + b x +c$, for values of $N$ between $3$ and
$35$, the fit yields $a = -0.272$, $b= -0.493$ and $c=0.244$ which is
in good agreement with that predicted here. However given the errors
for large $N$ and the fact that there are probably corrections of
$O(\ln(N))$, the numerical estimate for the exponent probably on has
about a 10 \% accuracy.
\begin{figure}
\centerline{\epsfxsize\columnwidth \epsfbox{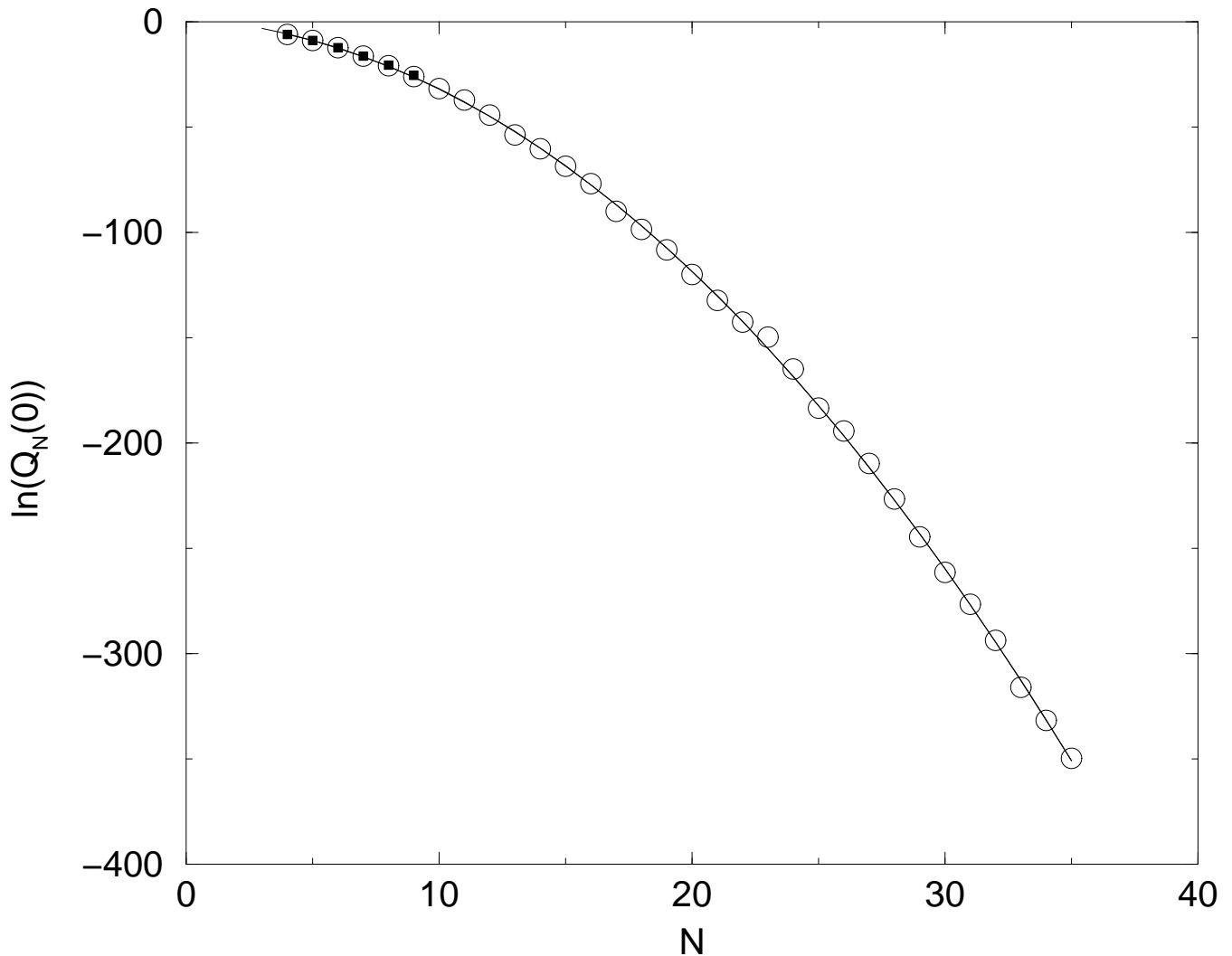}}
\caption{Monte Carlo computation of $\ln(Q_N(0))$ for the GOE (black
circles error at large $N$ indicated by the size of the circles) along with
quadratic fit (solid line). Shown as squares are the results obtained
by modified enumeration.}
\label{qplot}
\end{figure}

\begin{figure}
\centerline{\epsfxsize\columnwidth \epsfbox{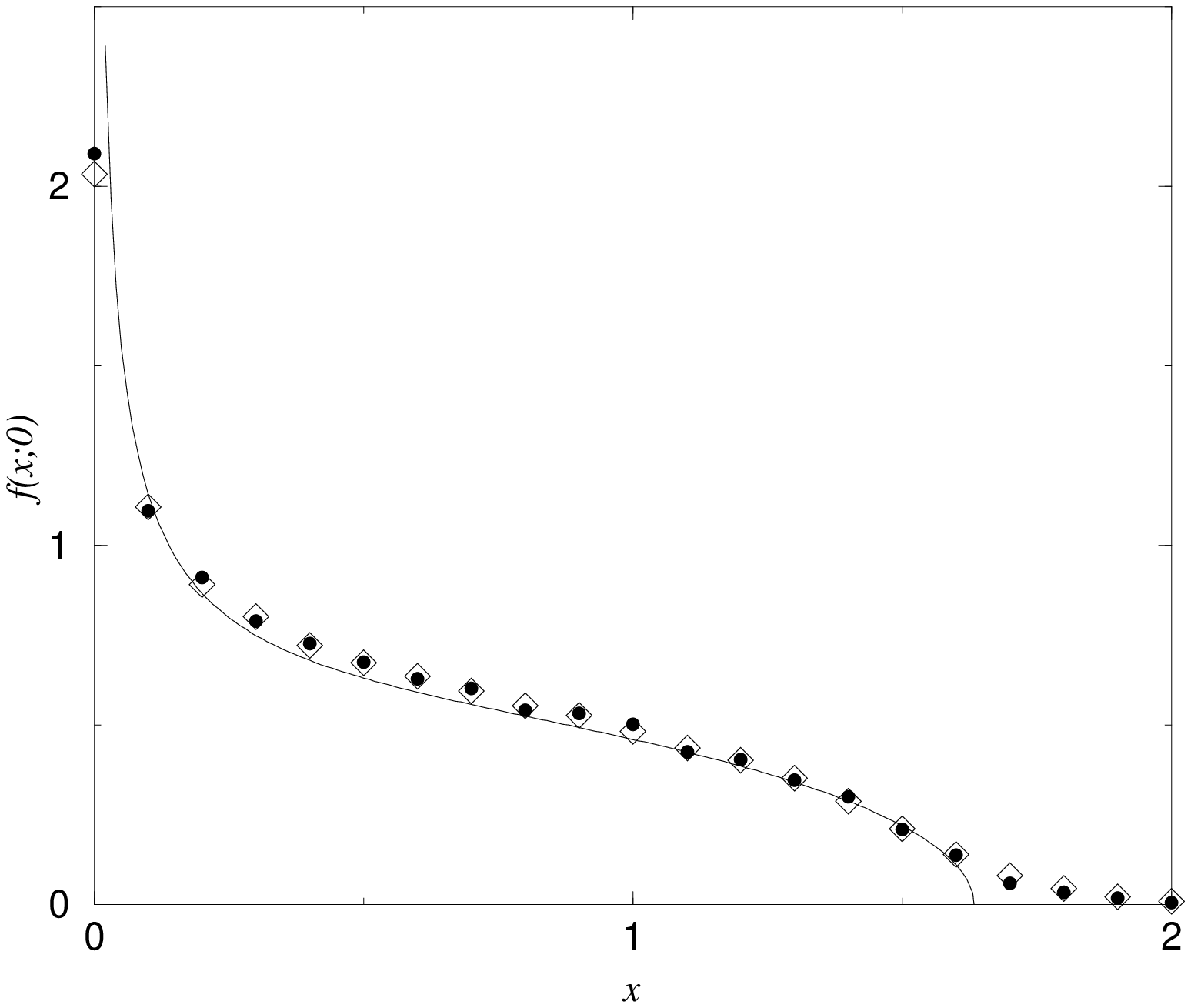}}
\caption{The analytical large $N$ formula for $f(x;0)$ with $z=0$ (solid line)
along with the numerically generated averaged histogram of $6\times 6$ (open
squares) and $7\times 7$ (solid circles)
Gaussian matrices with positive eigenvalues. The agreement is already 
good, the main difference occurring at the large $x$ tail.}
\label{f67plot}
\end{figure}

Also, by direct sampling over Gaussian matrices, one can numerically
evaluate the the rescaled density for states for matrices having only
positive eigenvalues. Because we use the direct sampling method we are
clearly restricted to small values of $N$, however in
Fig. (\ref{f67plot}) we show the analytical large $N$ result for $f$
with that computed numerically for matrices with $N=6$ and for $N=7$,
we see that despite the small value of $N$ the agreement is already
rather good, the main deviation being in the tails for large $\mu$.

\section{Conclusions}

In this paper we have shown how the Coulomb gas formulation of the
distribution of eigenvalues of $(N\times N)$ Gaussian random matrices
can be exploited to derive exact asymptotic results concerning the
extreme value statistics of their eigenvalues. Our main results are
summarized as follows.
\vskip 0.3cm

\noindent $(i)$ We have shown the probability $P_N$ that all
eigenvalues are positive (or negative) (or equivalently the
probability that $\lambda_{\rm min}\ge 0$ or $\lambda_{\rm max}\le 0$)
decays as $P_N\sim \exp[-\beta\, \theta(0)\, N^2]$ for large $N$ where
$\theta(0)=\ln(3)/4=0.274653\ldots$ and $\beta$ is the Dyson index.

\vskip 0.2cm

\noindent $(ii)$ More generally, we have computed the probability
${\rm Prob}[\lambda_{\rm max}\le t, N]$ that the maximal eigenvalue is
located deep within the Wigner sea region, far to the left from its
average value $\sqrt{2N}$, i.e., when $t\sim O(N^{1/2})\le \sqrt{2N}$.
This probability has the asymptotic form, $\sim \exp\left[-\beta\,
N^2\, \Phi\left( \frac{\sqrt{2N}-t}{\sqrt{N}}\right)\right]$ for large
$N$, where the large deviation function $\Phi(y)$ has been computed
exactly.

\vskip 0.2cm

\noindent $(iii)$ We have also computed the asymptotic joint probability
distribution of $\lambda_{\rm min}$ and $\lambda_{\rm max}$. 

\vskip 0.3cm

Our result in (ii), valid when $\sqrt{2N}-t \sim O(N^{1/2})$ (deep
inside the Wigner sea) is complimentary to the Tracy-Widom \cite{TW}
result that concerns the distribution of $\lambda_{\rm max}$ about its
mean value (near the edge) over a small range of width $\sim
N^{-1/6}$, i.e., for $\sqrt{2N}-t \sim O(N^{-1/6})$. We have
demonstrated explicitly how these two results match up smoothly as one
approaches from deep inside the Wigner sea to its right edge.

The key step in our method for computing the distribution of
$\lambda_{\rm min}$ (or $\lambda_{\rm max}$) in (ii) consists in using
a functional integral approach to study the Coulomb gas representation
of the problem and imposing a single hard wall constraint which
enforces the fact that no eigenvalues can be to the left (or right) of
a given point. For the computation of the joint distribution of
$\lambda_{\rm min}$ and $\lambda_{\rm max}$ in (iii), we needed to
confine the Coulomb gas within two hard walls.  In the limit of large
$N$, the functional integrals can be evaluated by the saddle point
method and the resulting integral equations for the saddle point
density can be solved explicitly using Tricomi's theorem
\cite{hhilbert}.

Our method is actually rather general and has already
been adapted to study the critical points of Gaussian random fields in
large dimensional spaces \cite{brde,FSW} and the extreme value statistics
of the maximum eigenvalue of Wishart random matrices \cite{vmb}.
One can possibly
find further applications in related statistical problems.
For instance the method is probably adaptable to study
the statistics of the index (the number of negative eigenvalues)
\cite{index} of random matrices and one could also study the extreme
value statistics of the minimal value of the modulus of the
eigenvalues by introducing the the appropriate constraint on the
density of eigenvalues in the functional integral formulation of the
problem.

Note that in this paper we were able to compute only the leading large $N$
behavior of the distribution of extreme eigenvalues. It would be interesting
to compute the sub-leading corrections to this leading behavior. Some recent
attempts have been made in this direction~\cite{OK}.
\vskip 1 truecm
 
{\bf Acknowledgments} We would like to thank V. Osipov for useful discussions
and for explaining his preliminary results with E. Kanzieper.


\begin{thebibliography}{10}
\bibitem{comment1} While in principle the $O(N)$ contributions could have possibly modified this
limiting behaviour, the agreement with the Tracy-Widom asymptotics shows that
clearly this is not the case.
\bibitem{comment2} There is  clearly a misprint in the sign of exponent of $N$ 
given for this fit in \cite{AE}.
\end{thebibliography}
\end{document}